\documentclass[letterpaper,english,aps,reprint,prb]{revtex4-1}
\usepackage[T1]{fontenc}
\usepackage[latin9]{inputenc}
\usepackage{amstext}
\usepackage{amssymb}
\usepackage{graphicx}

\makeatletter

\usepackage{amsmath}
\DeclareMathOperator*{\argmax}{argmax}
\DeclareMathOperator*{\argmin}{argmin}

\@ifundefined{showcaptionsetup}{}{%
 \PassOptionsToPackage{caption=false}{subfig}}
\usepackage{subfig}
\makeatother

\usepackage{babel}
\begin{document}
\global\long\def\hc{\text{h.c.}}%
 
\global\long\def\re{\text{Re}}%
 
\global\long\def\im{\text{Im}}%
 
\global\long\def\K{\mbox{ K}}%
 
\global\long\def\meV{\mbox{ meV}}%
 
\global\long\def\eV{\mbox{ eV}}%
 
\global\long\def\hf{\!\,_{2}F_{1}}%
 
\global\long\def\pr{\prime}%
 
\global\long\def\prpr{\prime\prime}%
 
\global\long\def\sgn{\text{sgn}}%
\global\long\def\amin{\argmin}%
\global\long\def\amax{\argmax}%

\title{Robust cluster expansion of multicomponent systems using structured
sparsity}
\author{Zhidong Leong}
\email{leong_zhidong@ihpc.a-star.edu.sg}

\author{Teck Leong Tan }
\email{Corresponding author: tantl@ihpc.a-star.edu.sg}

\affiliation{Institute of High Performance Computing, Agency for Science, Technology
and Research, Singapore 138632, Singapore}
\date{\today}
\begin{abstract}
Identifying a suitable set of descriptors for modeling physical systems
often utilizes either deep physical insights or statistical methods
such as compressed sensing. In statistical learning, a class of methods
known as structured sparsity regularization seeks to combine both
physics- and statistics-based approaches. Used in bioinformatics to
identify genes for the diagnosis of diseases, \textit{group lasso}
is a well-known example. Here in physics, we present group lasso as
an efficient method for obtaining robust cluster expansions (CE) of
multicomponent systems, a popular computational technique for modeling
such systems and studying their thermodynamic properties. Via convex
optimization, group lasso selects the most predictive set of atomic
clusters as descriptors in accordance with the physical insight that
if a cluster is selected, so should its subclusters. These selection
rules avoid spuriously large fitting parameters by redistributing
them among lower order terms, resulting in more physical, accurate,
and robust CEs. We showcase these features of group lasso using the
CE of bcc ternary alloy Mo-V-Nb. These results are timely given the
growing interests in applying CE to increasingly complex systems,
which demand a more reliable machine learning methodology to handle
the larger parameter space. 
\end{abstract}
\maketitle

\section{Introduction \label{sec:Introduction}}

Model building in physics requires both physical insights and statistics.
In the cluster expansion (CE) of multicomponent systems \citep{Sanchez1984},
physical insights prescribe that the energies of atomic configurations
obey a generalized Ising-like Hamiltonian. The energy $E\left(\sigma\right)$
of an atomic structure $\sigma$ can be expanded in terms of atomic
clusters $\alpha$, where the cluster correlation functions $\Phi_{\alpha}\left(\sigma\right)$
serve as the basis set and the effective cluster interactions (ECIs)
$V_{\alpha}$ as the coefficients: 
\begin{eqnarray}
E\left(\sigma\right) & = & \sum_{\alpha}\Phi_{\alpha}\left(\sigma\right)V_{\alpha}.\label{eq:clus-expansion}
\end{eqnarray}
Statistically optimal values of the ECIs could be obtained via fitting
to Eq. \ref{eq:clus-expansion} the energies of a training set of
structures, usually calculated from first principles. When appropriately
truncated, the CE is an accurate model for efficiently predicting
the energies \citep{Blum2004,Tan2012,Ng2013,Wrobel2015,Maisel2016}
or associated properties \citep{Ferreira1991,VanderVen2001,Chan2010,Maisel2012,Wang2012,Fernandez-Caballero2017}
of different atomic configurations.

However, selecting the appropriate set of atomic clusters as descriptors
is challenging: selections based on physical intuition are not robust,
while those based on statistics are not physical. Initially, CE was
largely applied to binary alloys \citep{Connolly1983,Lu1991,Wolverton1992,Zunger1994,Garbulsky1994,Fontaine1994,Lu1995,Garbulsky1995,Wolverton1995,Ozolins1998,Kohan1998,Muller2001,Zunger2002,vandeWalle2002c,Blum2004}.
Thereafter, it has been applied to more complex systems, including
ternary to quinary alloys \citep{Wrobel2015,Maisel2016,Ji2017,Feng2017,Nguyen2017,Fernandez-Caballero2017},
semiconductors \citep{Ferreira1991,Burton2011}, battery materials
\citep{VanderVen2004,Persson2010}, clathrates \citep{Angqvist2016,Angqvist2017},
magnetic alloys \citep{Drautz2004,Drautz2005,Lavrentiev2010}, and
nanoscale alloys \citep{Wang2012,Tan2012,Kang2013,Ng2013,Cao2015,Cao2016,Tan2017,Cao2018}.
In complex systems, the reduced symmetry increases the number of symmetrically
distinct clusters, exacerbating the cluster selection problem. With
growing enthusiasm in applying CE to higher component systems, such
as high-entropy alloys \citep{Fernandez-Caballero2017}, it is timely
to introduce an improved machine-learning procedure for creating reliable
CEs with physically meaningful and robust ECIs.

Currently, there are two prevalent approaches for cluster selection.
The first emphasizes using physical insights, such as via specific
priors in the Bayesian framework \citep{Mueller2009} or via selection
rules to incorporate smaller clusters before larger ones \citep{vandeWalle2002a,Zarkevich2004}.
The second approach espouses using sparsity-driven regularization
such as compressed sensing \citep{Hart2005,Drautz2006,Nelson2013,Nelson2013a,Maisel2016,Angqvist2016,Angqvist2017}.
Fundamentally, CE is a standard linear regression problem $y=X\beta$---the
response $y_{i}$ is the first-principles energy of the $i$th structure
in the training set $\left\{ \sigma\right\} $, the coefficient $\beta_{j}$
is the ECI of the $j$th cluster, and the component $x_{ij}$ of the
design matrix $X$ is the correlation function $\Phi_{j}\left(\sigma_{i}\right)$
of structure $i$ with respect to cluster $j$. Typically, the optimal
$\hat{\beta}$ is given by the regularized least-squares solution
\begin{eqnarray}
\hat{\beta} & = & \amin_{\beta}\left\Vert y-X\beta\right\Vert _{2}^{2}+g\left(\beta\right),\label{eq:reg-least-sq}
\end{eqnarray}
where the $\ell_{p}$-norm is defined by $\left\Vert z\right\Vert _{p}=\left(\sum_{i}\left|z_{i}\right|^{p}\right)^{1/p}.$
The penalty function $g\left(\beta\right)$ constrains $\beta$ to
reduce overfitting and is key to high prediction accuracy for structures
outside the training set. In compressed sensing \citep{Candes2006,Candes2008},
the least absolute shrinkage and selection operator (lasso) $g\left(\beta\right)\propto\left\Vert \beta\right\Vert _{1}$
selects atomic clusters by favoring parsimonious models \citep{Nelson2013,Nelson2013a};
such models are more interpretable and simpler for quick computation,
for example, in Monte-Carlo simulations.

In this paper, we present \textit{group lasso} regularization \citep{Yuan2006}
as an efficient method for obtaining reliable CEs of multicomponent
systems. As an example of structured sparsity in machine learning,
group lasso combines sparsity-driven regularization with physical
insights to select atomic clusters as descriptors. We show that even
with the large parameter space of ternary alloys and beyond, the resulting
truncated CE remains sparse and robust with interpretable ECIs. With
a specially constructed convex penalty $g\left(\beta\right)$, group
lasso imposes the physical insight that a cluster is selected only
after all its subclusters. These selection rules avoid spuriously
large fitting parameters by redistributing them among lower order
terms, resulting in more physical, accurate, and robust CEs. We will
demonstrate these features of group lasso via the CE of ternary bcc
alloy Mo-V-Nb.

\section{Methods}

\subsection{Group lasso \label{subsec:group-lasso}}

\begin{figure}
\subfloat[\label{fig:lasso-constraint-plot-3d}]{\includegraphics[scale=0.25]{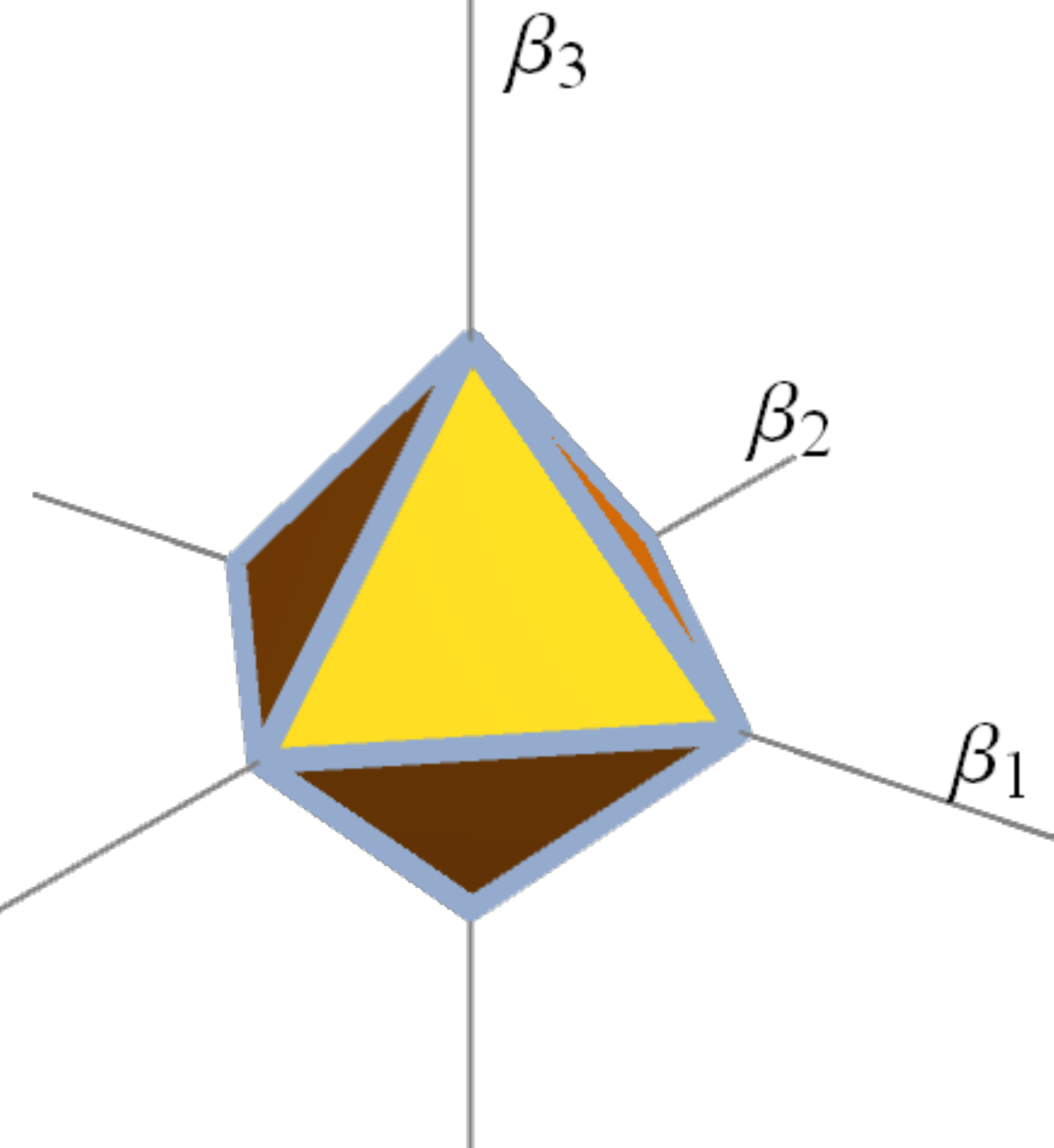}

}\subfloat[\label{fig:glasso-constraint-plot-3d}]{\includegraphics[scale=0.25]{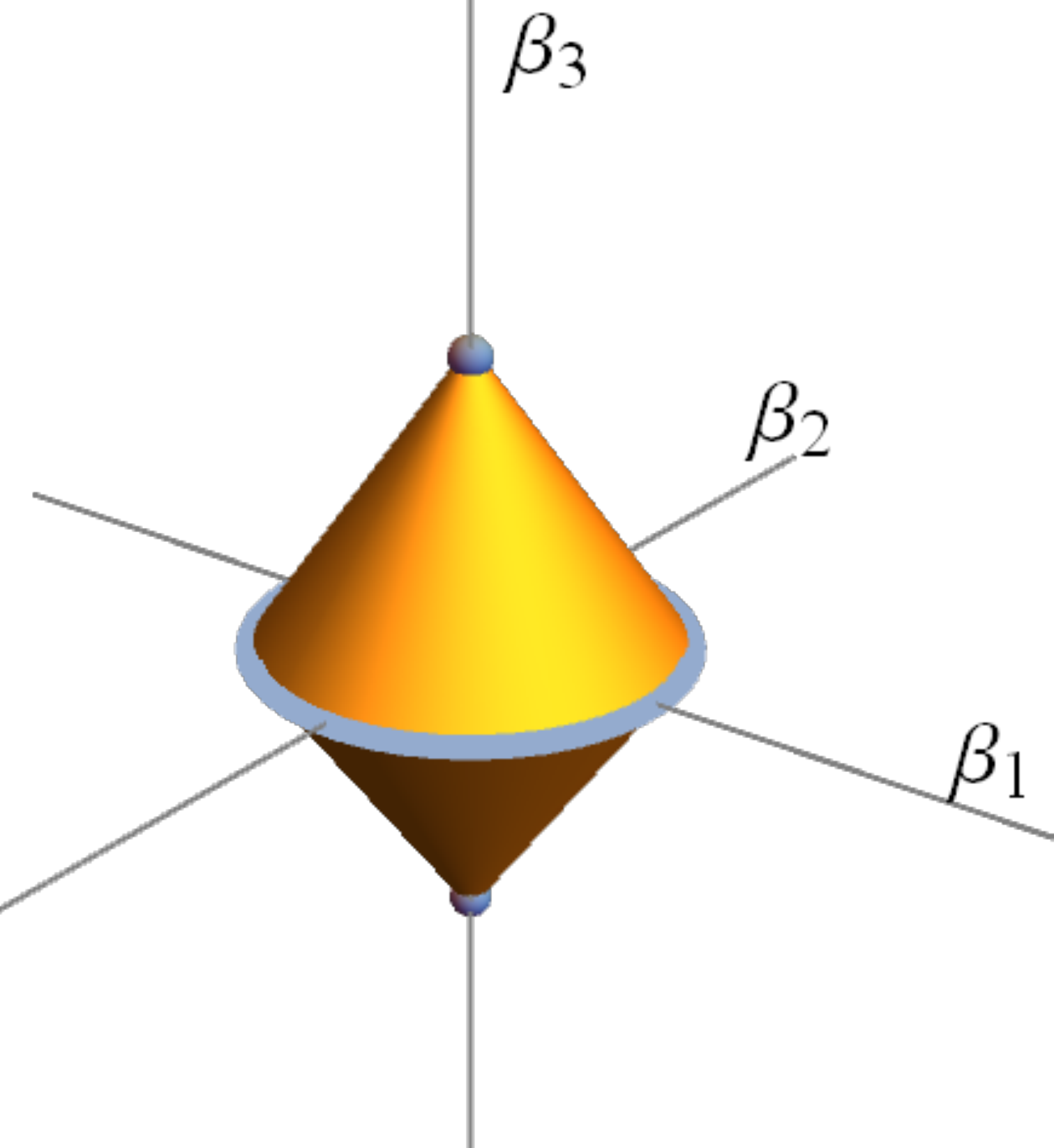}

}

\caption{The constraints on $\left\{ \beta_{1},\beta_{2},\beta_{3}\right\} $
in (a) lasso and (b) group lasso regularizations. The corners/edges
(in light blue) of these constraints correspond to sparse solutions.
In (b), coefficients $\beta_{1}$ and $\beta_{2}$ are grouped, while
$\beta_{3}$ remains a singleton. This grouping favors solutions with
$\beta_{1},\beta_{2}$ either both zero or both nonzero.}
\end{figure}

Group lasso is an extension of the well-known lasso regularization
\citep{Tibshirani1996,Hastie2015}. Using the nonanalyticity of the
penalty functions, both methods favor sparse solutions to the linear
regression problem $y=X\beta$. For example, the lasso penalty is
$g\left(\beta\right)=\lambda\left\Vert \beta\right\Vert _{1}$ with
hyperparameter $\lambda$, which has been studied in the context of
compressed sensing CE \citep{Nelson2013,Nelson2013a}. In this case,
the sparsity of the regularized solution from Eq. \ref{eq:reg-least-sq}
can be understood in the dual picture 
\begin{eqnarray}
\hat{\beta} & = & \amin_{\beta}\left\Vert y-X\beta\right\Vert _{2}^{2},\text{ with }\left\Vert \beta\right\Vert _{1}<\tau,
\end{eqnarray}
where $\tau$ is inversely related to $\lambda$. Fig. \ref{fig:lasso-constraint-plot-3d}
illustrates the constraint $\left\Vert \beta\right\Vert _{1}<\tau$
for $\beta\in\mathbb{R}^{3}$. This constraint shrinks the least-squares
solution to one that tends to lie on the corners/edges of the constraint
highlighted in Fig. \ref{fig:lasso-constraint-plot-3d}. The resulting
regularized solution is therefore sparse with some $\hat{\beta}_{i}$
vanishing. 

In conventional lasso, the sparse solution is determined from a statistical
fit, with little room for incorporating pertinent physical insights.
In contrast, group lasso seeks a more physically meaningful solution
by ensuring that physically-related coefficients are either all zero
or all nonzero together as a group. For example, when applied to gene
expression data for the diagnosis of diseases in bioinformatics, group
lasso ensures that genes with coordinated functions are either all
excluded or all included in the model \citep{Ma2007}. For CE, we
will use group lasso to impose physical cluster selection rules.

In group lasso, the coefficients $\beta$ are partitioned into $J$
groups $\theta_{1},\ldots,\theta_{J}$, where $\theta_{j}\in\mathbb{R}^{p_{j}}$
is a group of $p_{j}$ coefficients. Let $Z_{j}$ be the matrix formed
by the columns of $X$ corresponding to the group $\theta_{j}$. Then,
the regularized solution is

\begin{eqnarray}
\hat{\beta} & = & \amin_{\beta}\frac{1}{2}\left\Vert y-\sum_{j=1}^{J}Z_{j}\theta_{j}\right\Vert _{2}^{2}+\lambda\sum_{j=1}^{J}\sqrt{p_{j}}\left\Vert \theta_{j}\right\Vert _{2},\label{eq:glasso-solution}
\end{eqnarray}
with hyperparameter $\lambda$. Notice that unlike in the least-squares
term, the $\ell_{2}$-norm in the penalty is not squared and is therefore
nonanalytic. It is this nonanalyticity that imposes sparsity. 

In the dual picture, the unregularized least-squares solution is now
constrained by 
\begin{eqnarray}
\sum_{j=1}^{J}\sqrt{p_{j}}\left\Vert \theta_{j}\right\Vert _{2} & < & \tau.
\end{eqnarray}
Fig. \ref{fig:glasso-constraint-plot-3d} illustrates this group-lasso
constraint for the case with three coefficients and the groups $\theta_{1}=\left(\beta_{1},\beta_{2}\right)$
and $\theta_{2}=\beta_{3}$. In this case, Eq. \ref{eq:glasso-solution}
simplifies to 
\begin{eqnarray}
\hat{\beta} & = & \amin_{\beta}\frac{1}{2}\left\Vert y-X\beta\right\Vert _{2}^{2}+\lambda\left(\sqrt{2}\sqrt{\beta_{1}^{2}+\beta_{2}^{2}}+\left|\beta_{3}\right|\right).
\end{eqnarray}
Compared to the lasso case in Fig. \ref{fig:lasso-constraint-plot-3d},
sharp corners/edges (representing sparse solutions) are now at $\beta_{1},\beta_{2}\neq0,\beta_{3}=0$
and $\beta_{1}=\beta_{2}=0,\beta_{3}\neq0$. Group lasso thus favors
solutions with $\beta_{1},\beta_{2}$ either both zero or both nonzero.
In general, coefficients in the same group $\theta_{j}$ are either
all zero or all nonzero. 

When each group in Eq. \ref{eq:glasso-solution} is a singleton, that
is $p_{j}=1$ for all $j$, the regularized solution reduces to that
of lasso

\begin{eqnarray}
\hat{\beta}_{\text{lasso}} & = & \amin_{\beta}\frac{1}{2}\left\Vert y-X\beta\right\Vert _{2}^{2}+\lambda\left\Vert \beta\right\Vert _{1}.\label{eq:lasso-solution}
\end{eqnarray}
We will use this to benchmark the performance of group lasso. Since
the penalty terms for both lasso and group lasso are convex, the regularized
solutions can be efficiently obtained by convex optimization. Note
that the weights $\sqrt{p_{j}}$ in the penalty term of Eq. \ref{eq:glasso-solution}
ensure that groups of different sizes are penalized equally. Without
these weights, a group with many coefficients will unfairly dominate
the penalty term. We next discuss the cluster selection rules we wish
to impose using group lasso.

\subsection{Hierarchical cluster selection rules \label{subsec:-selection-rules}}

In CE, the energy of an atomic configuration is expanded in terms
of the atomic clusters and their associated ECIs. In general, since
a cluster $b$ is a higher order correction to its subcluster $a\subset b$,
the ECI $\beta_{b}\neq0$ only if the subcluster ECI $\beta_{a}\neq0$.
I.e., a CE should include a cluster only if all its subclusters are
also included. This is the hierarchical cluster selection rule we
adopt here. Similar rules have been used for the CEs of binary systems
\citep{vandeWalle2002a,Zarkevich2004,Sluiter2005,Zarkevich2008,Mueller2009,Tan2012,Ng2013,Tan2017}.
Here, we extend such rules to alloy systems with more components.

\begin{figure}
\subfloat[\label{fig:three-clusters}]{\includegraphics[scale=0.25]{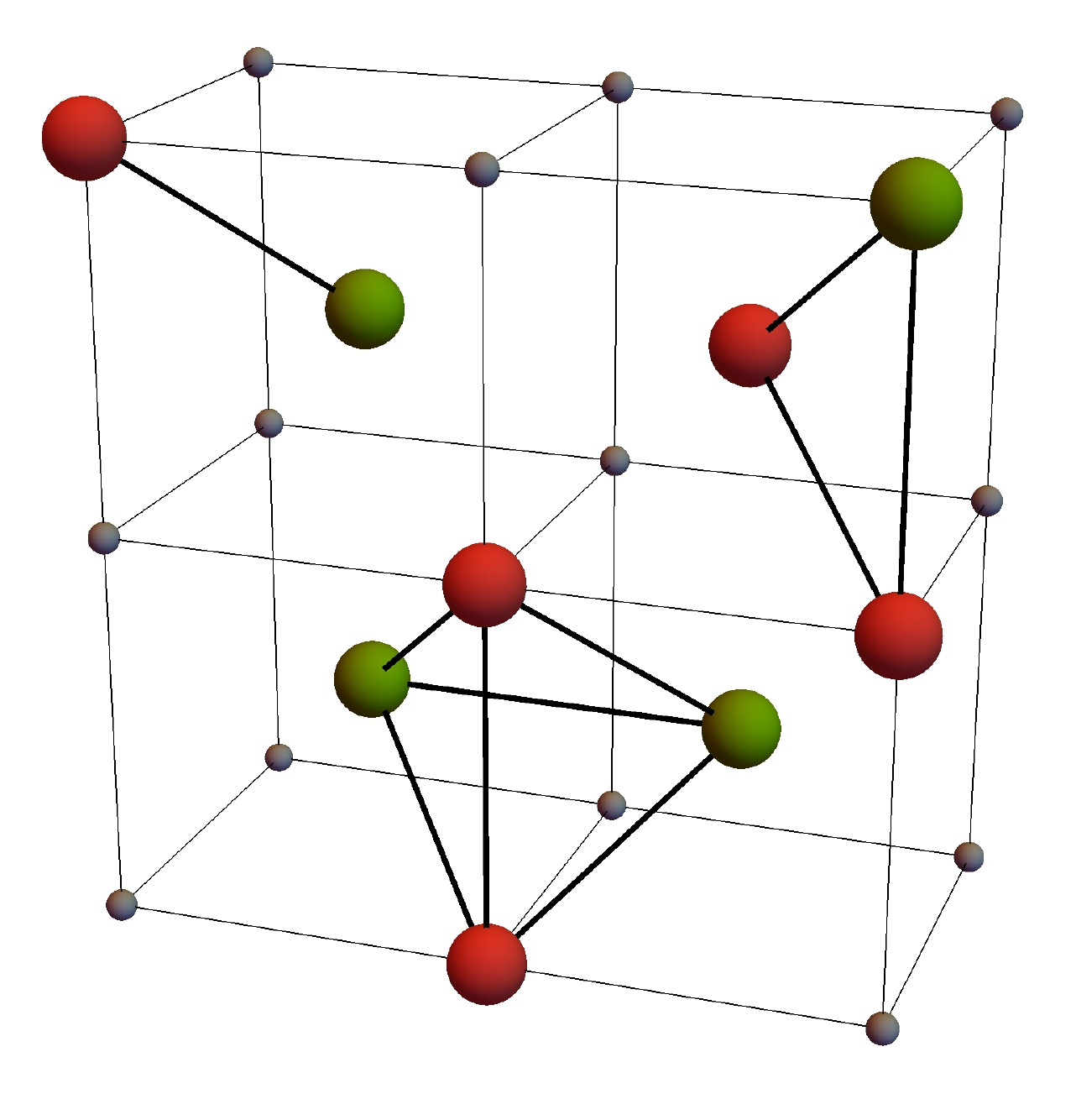}

}

\subfloat[\label{fig:subset-relations-graph}]{\includegraphics[scale=0.5]{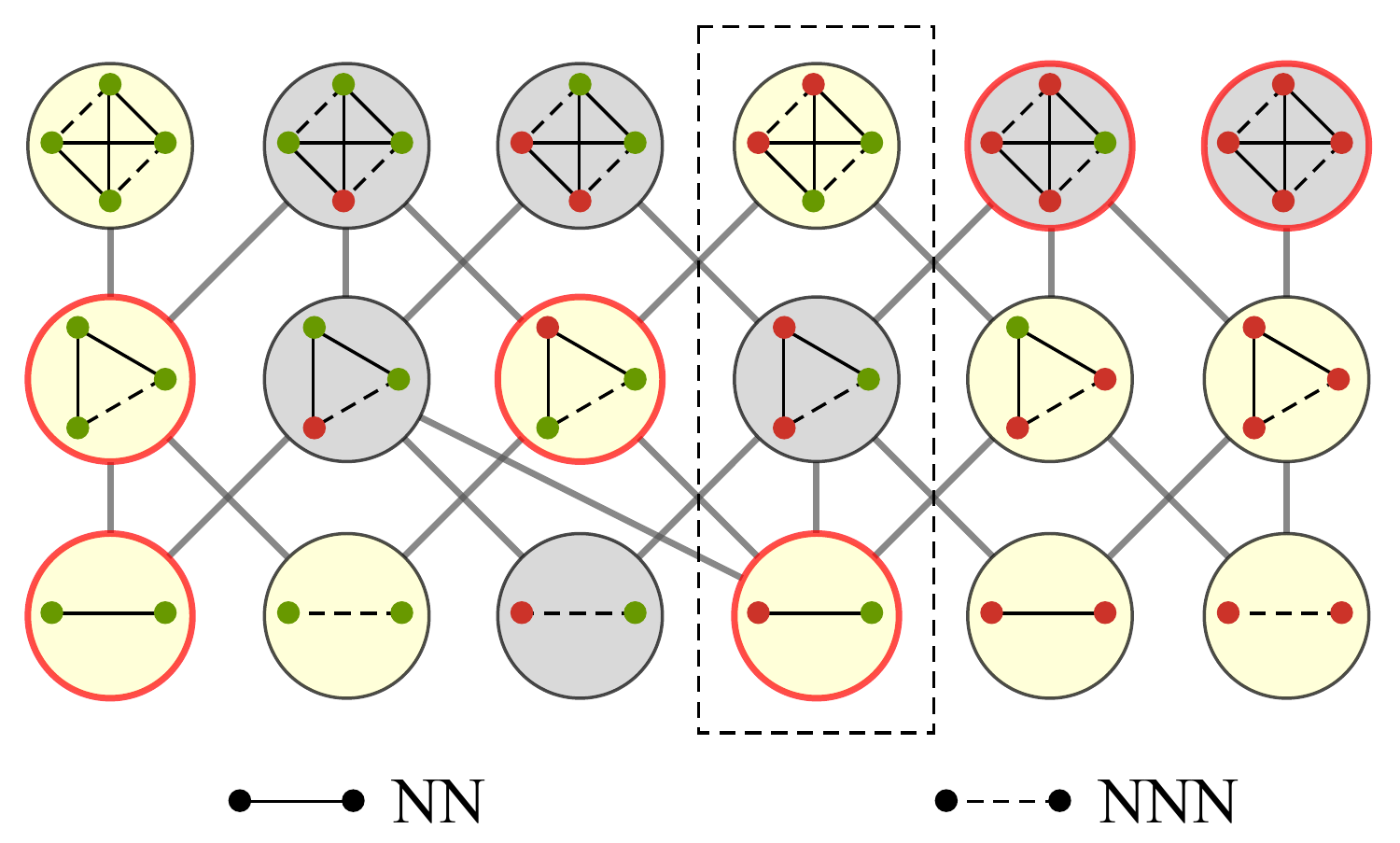}

}

\caption{Atomic clusters of a bcc ternary system, with atomic species distinguished
by colors. (a) Examples of the smallest pair, triplet, and quadruplet.
(b) A graphical representation of the subcluster relations. Each bubble
contains a cluster shown as a 2D schematic, with lines connecting
it to all its subclusters with one fewer atom. The clusters in the
dashed box correspond to those in (a). The highlighted vertices form
a set of clusters obeying the hierarchical selection rules, while
those with a red border do not.}
\end{figure}

Without vacancies, an $m$-component system requires the tracking
of $m-1$ independent atomic species. For $m\geq3$, the key distinction
from binaries is that for a given cluster, multiple decorations (of
independent atomic species) need to be accounted for when considering
subcluster relations. 
For a given independent decoration, the correlation function in Eq. \ref{eq:clus-expansion} is defined as the number of clusters present in the atomic structure.
For example, Fig. \ref{fig:three-clusters}
shows three decorated clusters of a ternary system on a bcc lattice.
The pair $a$, triplet $b$, and quadruplet $c$ are related by $a\subset b$,
$a\subset c$ and $b\not\subset c$. These relations are represented
graphically in Fig. \ref{fig:subset-relations-graph}, where each
bubble contains a cluster (shown as a 2D schematic) with lines connecting
it to its subclusters with one fewer atom. The three clusters in Fig.
\ref{fig:three-clusters} correspond to those in the dashed box in
Fig. \ref{fig:subset-relations-graph}. The set of highlighted clusters
(bubbles with yellow background) is an example satisfying the hierarchical
cluster selection rules, while the set with red borders does not.
Our work aims to use group lasso to obtain cluster sets that obey
the hierarchical rules.

\subsection{Cluster selection with group lasso}

Imposing the cluster selection rules using group lasso is a subtle
but important point. This is because the hierarchical rules require
overlapping groups of ECIs, which are incompatible with how group
lasso is formulated in Sec. \ref{subsec:group-lasso}. The solution
is to use a variant of group lasso known as \textit{overlap group
lasso} \citep{Jacob2009}.

To show how this variant of group lasso can impose the cluster selection
rules, we consider just two clusters $c_{1}\subset c_{2}$ and the
corresponding ECIs $\beta_{1}$ and $\beta_{2}$. To have $\beta_{2}\neq0$
imply $\beta_{1}\neq0$ (as per the selection rules), we first write
$\beta_{1}=\theta_{11}+\theta_{21}$ and $\beta_{2}=\theta_{22}$.
Then, grouping together $\theta_{21}$ and $\theta_{22}$, we apply
group lasso using Eq. \ref{eq:glasso-solution} to find the optimal
$\theta_{11},\theta_{21}$, and $\theta_{22}$: 
\begin{eqnarray}
\hat{\theta} & = & \amin_{\theta}\frac{1}{2}\left\Vert y-x_{1}\left(\theta_{11}+\theta_{21}\right)-x_{2}\theta_{22}\right\Vert _{2}^{2}\nonumber \\
 &  & +\lambda\left(\left|\theta_{11}\right|+\sqrt{2}\sqrt{\theta_{21}^{2}+\theta_{22}^{2}}\right).\label{eq:oglasso-solution-2d}
\end{eqnarray}
As discussed, the form of group lasso's penalty ensures that $\theta_{21}$
and $\theta_{22}$ are either both zero or both nonzero. Consequently,
$\beta_{2}\neq0$ implies that $\beta_{1}\neq0$ (almost surely),
but we can still have $\beta_{2}=0$ with $\beta_{1}\neq0$. This
is precisely the selection rule corresponding to the subcluster relation
$c_{1}\subset c_{2}$. 

For a general set of $p$ clusters $\left\{ c_{1},\ldots,c_{p}\right\} $,
group lasso can similarly impose the selection rules. First, we write
the ECIs $\beta=\left(\beta_{1},\ldots,\beta_{p}\right)^{T}$ as a
sum of $p$ groups of coefficients: $\beta=\sum_{j=1}^{p}\nu_{j}$
where $\nu_{j}\in\mathbb{R}^{p}$ is a vector constrained to be zero
everywhere except in positions corresponding to $c_{j}$ and its subclusters.
That is, we fix $v_{j,k}=0$ for all $k$ such that $c_{k}\not\subseteq c_{j}$.
Then, the group lasso solution for the unconstrained components is
analogous to Eq. \ref{eq:oglasso-solution-2d}:
\begin{eqnarray}
\hat{\nu} & = & \amin_{\nu}\frac{1}{2}\left\Vert y-X\sum_{j=1}^{p}\nu_{j}\right\Vert _{2}^{2}+\lambda\sum_{j=1}^{p}\sqrt{p_{j}}\left\Vert \nu_{j}\right\Vert _{2},\label{eq:oglasso-solution}
\end{eqnarray}
where $p_{j}$ is the number of subclusters of $c_{j}$ (including
$c_{j}$ itself). That is, $p_{j}$ is the number of unconstrained
components in $\nu_{j}$.

Here, we verify that Eq. \ref{eq:oglasso-solution} works as intended:
the selection of a cluster $c_{j}$ should imply the selection of
its subcluster $c_{l}\subset c_{j}$. Given $\beta_{j}\neq0$, we
have $\nu_{k,j}\neq0$ for some $k$ such that $c_{j}\subseteq c_{k}$.
Then, for a subcluster $c_{l}\subset c_{j}$ (and hence $c_{l}\subset c_{k}$),
the $\left\Vert v_{k}\right\Vert _{2}$ term in the penalty ensures
that $\nu_{k,l}\neq0$. Consequently, $\beta_{l}\neq0$, as required.

\section{Results}

We showcase the features of group lasso via the CE of bcc ternary
alloy Mo-V-Nb, whose constituent elements are well-known refractory
metals. Previously, CE has been used to study the ground states of
binary alloys V-Nb \citep{Ravi2012} and Mo-Nb \citep{Blum2005,Huhn2013}.
Here, we benchmark the performance of group lasso (Eq. \ref{eq:oglasso-solution})
against lasso (Eq. \ref{eq:lasso-solution}). The former method imposes
the hierarchical cluster selection rules, while the latter performs
regularization based just on statistics. The value of the hyperparameter
$\lambda$ in each method is fixed by cross-validation (CV). Our training
structures have small unit cells with up to six atoms. We use 239
clusters consisting of pairs, triplets, ..., and six-body clusters,
with 1654 cluster selection rules. As we will see, group lasso tends
to produce CEs that are more physical, accurate, and robust than those
from lasso. The appendix contains further technical details about
our implementation. 

\paragraph*{Physicalness:}

\begin{figure}
\includegraphics[scale=0.5]{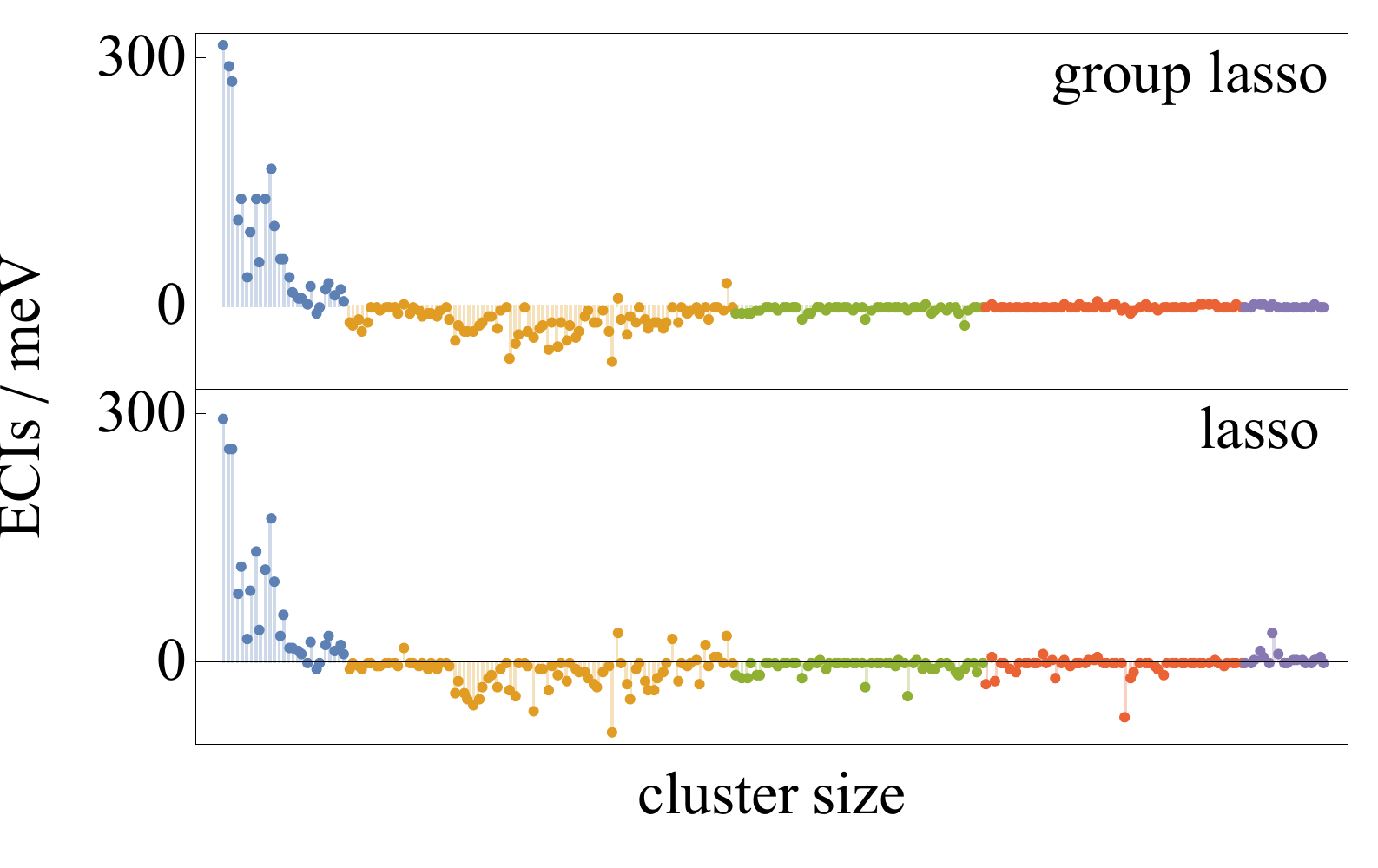}

\caption{The values of $239$ ECIs based on $800$ training structures. Pairs,
triplets, quadruplets, 5-bodies, and 6-bodies are colored blue, orange,
green, red, and purple, respectively. The ECIs from group lasso are
well-behaved---larger clusters generally have smaller ECIs---while
for lasso, several isolated spikes corresponding to large ECIs are
observed among the higher-order clusters (quadruplets and beyond).
\label{fig:glasso-ecis}}
\end{figure}

Fig. \ref{fig:glasso-ecis} shows the values of the ECIs based on
$800$ training structures. The group lasso ECIs, by construction,
obey all the cluster selection rules, and they satisfy the physical
intuition that ECIs generally weaken with increasing cluster size.
This behavior suggests that the CE is converging, given our initial
pool of clusters. In contrast, the lasso ECIs obey only $\sim87\%$
of the rules, and numerous large clusters have abnormally large ECIs.
Therefore, via the selection rules, group lasso redistributes these
spurious spikes in lasso among lower-order terms. While this redistribution
decreases sparsity (205 nonzero ECIs for group lasso vs 180 for lasso),
CEs from group lasso have more physical trends in the ECIs than from
lasso. These general behaviors are observed regardless of the training
set choices.

\paragraph*{Accuracy: }

\begin{figure}
\includegraphics[scale=0.5]{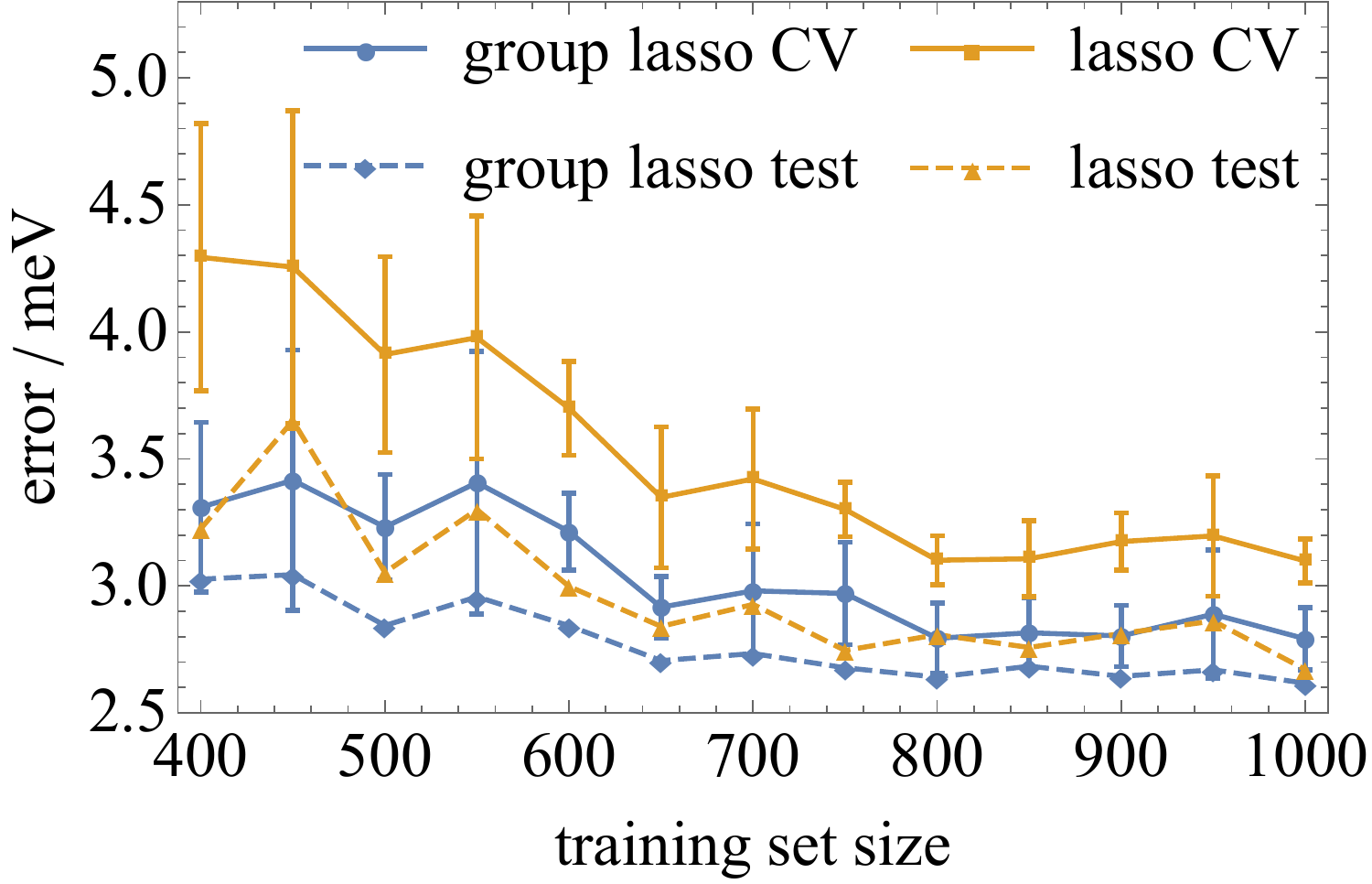}

\caption{Five-fold cross-validation (CV) scores and test errors for group lasso
and lasso versus training set size. The errors of group lasso are
consistently lower than lasso's. The error bars for the CV scores
correspond to one standard deviation among the five folds. \label{fig:test-error-vs-ntrain}}
\end{figure}

In addition to the training structures, we also have $500$ test structures
with large $16$-atom unit cells not used for training. For both lasso
and group lasso, Fig. \ref{fig:test-error-vs-ntrain} shows the CV
scores and test errors decreasing as the number of training structures
increases, signifying the convergence of the CEs. For either method,
the CV scores and test errors are comparable. These observations imply
that the lasso class of methods are able to distill the essential
physics from training with just small structures, reliably predicting
the energies of larger structures not in the training set. This is
advantageous for ternary alloys and beyond, because of the huge number
of large structures in these systems. For all training set sizes,
group lasso is consistently more accurate than lasso (smaller CV scores
and test errors). Therefore, the incorporation of physical hierarchy
improves not only the physical interpretability of the ECIs but also
the predictive capability of the CE. Group lasso reduces overfitting
by redistributing the contributions from unphysical spikes in lasso's
ECIs among numerous smaller clusters that are more important. 

\paragraph*{Robustness:}

\begin{figure}
\includegraphics[scale=0.5]{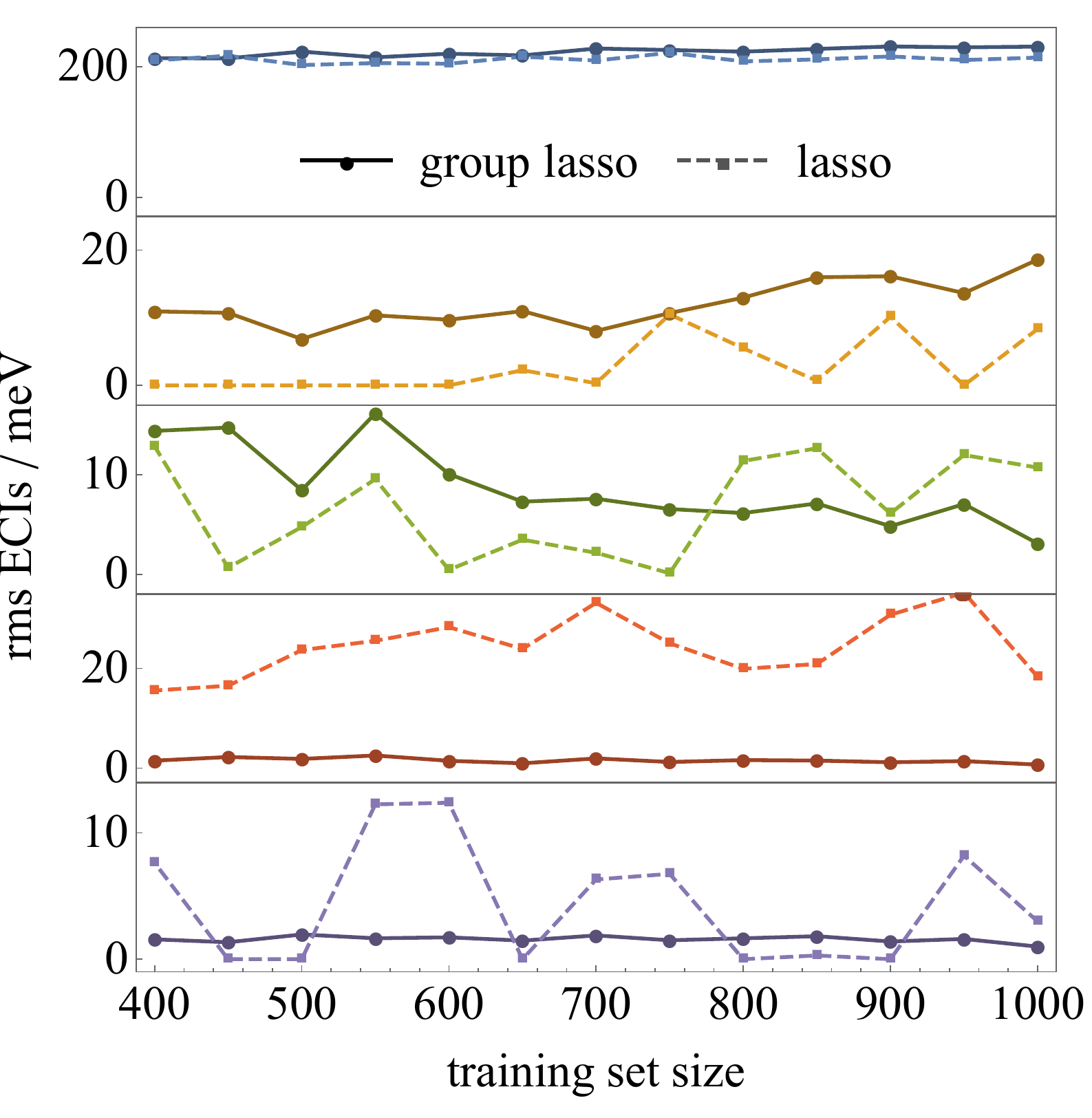}

\caption{The root-mean-square ECIs with respect to the number of training structures
for different category of clusters, namely, from top to bottom, pairs,
triplets, quadruplets, 5-bodies, and 6-bodies. For higher order clusters,
the ECIs of group lasso tend to fluctuate much less than those of
lasso. \label{fig:mean-eci-vs-ntrain}}
\end{figure}

The ECIs of a robust CE should converge towards the true physical
values when more training structures are used. As such, a lack of
robustness is signified by ECIs wildly fluctuating with respect to
the size of the training set. The degree of fluctuations can be concisely
illustrated using the root-mean-square (rms) of the ECIs in each cluster
category (pairs, triplets, $\ldots$, and six-bodies). Fig. \ref{fig:mean-eci-vs-ntrain}
shows that the five rms ECIs from group lasso are largely stable with
respect to the number of training structures. However, the ECIs from
lasso tend to vary wildly for the higher order clusters. This distinction
shows that group lasso produces CEs that are more robust; the ECIs
are more physically interpretable for group lasso (especially for
higher order clusters), as they tend to fluctuate less with different
training sets.

\section{Discussions and conclusion}

As mentioned in Sec. \ref{subsec:-selection-rules}, similar hierarchical
cluster selection rules have been used for CE \citep{vandeWalle2002a,Zarkevich2004,Sluiter2005,Zarkevich2008,Mueller2009,Tan2012,Ng2013,Tan2017}.
Compared to previous works, the combination of these rules with sparsity-driven
regularization in our work leads to more robust ECIs. This is because
regularization shrinks the values of the selected ECIs to avoid spuriously
large terms. Furthermore, since previous methods involve evaluating
different combinations of clusters separately to find the optimal
one, these methods become less computationally feasible for ternary
systems and beyond, where many more combinations of clusters need
to be explored. This is so unless the search space is shrunk by imposing
additional selection criteria, for example, if an $n$-body cluster
is included, then all $n$-body clusters of smaller spatial extent
are also included \citep{vandeWalle2002a,Zarkevich2004}. We do not
impose these additional criteria in our work; they might be too restrictive
for ternaries and beyond because, for example, the inclusion of A-B
pairs up to a certain spatial extent should not impact the spatial
extent of pairs for other decorations (i.e., B-C, A-C).

In Ref. \citep{Sluiter2005}, the authors studied the invariance of CE under linear
transformations of the site occupation variables. The authors showed that invariance is preserved only when 
the hierarchical cluster selection rules are obeyed. We emphasize that our group lasso implementation obeys the hierarchical rules, whereas standard lasso does not. 
Hence, our work presents a way for preserving the invariance of CE.

In conclusion, we presented \textit{group lasso} \citep{Yuan2006}
as an efficient method for producing reliable CEs of multicomponent
alloys, resulting in accurate and robust surrogate models for predicting
thermodynamic properties. A type of structured sparsity regularization,
group lasso combines statistical learning with physical insights to
select atomic clusters as descriptors for the CE model. Via convex
optimization, group lasso imposes the cluster selection rules that
a cluster is selected only after all its subclusters. These rules
avoid spuriously large fitting parameters by redistributing them among
numerous lower order terms, resulting in more physical, accurate,
and robust CEs. These results are timely given the growing interests
in applying CE to increasingly complex systems, where the larger parameter
space demands a more reliable machine learning methodology to construct
robust models. Furthermore, this work should inspire applying structured
sparsity in modeling other physical systems.
\begin{acknowledgments}
The authors thank the Advanced Manufacturing and Engineering Young
Individual Research Grant (AME YIRG) of Agency for Science, Technology
and Research (A{*}STAR) (A1884c0016) for financial support. The DFT
computations in this article were performed on the resources of the
National Supercomputing Centre, Singapore (https://www.nscc.sg).
\end{acknowledgments}

\appendix*

\section*{Appendix}

In this appendix, we present the technical details about our implementation
of cluster expansion (CE) and group lasso.

\subsection{First-principles calculations}

The energies of the training and test structures are calculated based
on density functional theory (DFT) with the Vienna Ab initio Simulation
Package (VASP) \citep{Kresse1996,Kresse1996a}. We use the Perdew,
Burke, and Ernzerhof exchange correlation based on the generalized
gradient approximation \citep{Perdew1996,Perdew1997}. The PAW potentials
are used with the outer $p$ semi-core states included in the valence
states \citep{Blochl1994,Kresse1999}. Plane-wave cutoffs are set
to 520 eV and all atomic coordinates (including lattice vectors) were
fully relaxed until the calculated Hellmann-Feynman force on each
atom was less than $0.015\eV/\text{\AA}$. Calculations are non spin-polarized
as Mo, Nb, and V are not known to be strongly magnetic. The $k$-point
mesh is generated using a Gamma grid and density of $200\ \text{\AA}^{-3}$.

\subsection{Normalization choice for cluster correlations \label{subsec:cluster-correlation}}

The general expression for CE given by 

\begin{eqnarray}
E\left(\sigma\right) & = & \sum_{\alpha}\Phi_{\alpha}\left(\sigma\right)V_{\alpha}.\label{eq:clus-expansion-1}
\end{eqnarray}
can be rewritten to account for the degeneracy of the clusters in
a specific lattice \citep{Zarkevich2007}. For any rescaling factor
$\eta_{\alpha}>0$, Eq. \ref{eq:clus-expansion-1} is invariant under
the transformation $\Phi_{\alpha}\left(\sigma\right)\rightarrow\Phi_{\alpha}\left(\sigma\right)\eta_{\alpha}$
and $V_{\alpha}\rightarrow V_{\alpha}/\eta_{\alpha}$. The choice
of $\eta_{\alpha}$ depends on whether degeneracy factors are subsumed
into $\Phi_{\alpha}\left(\sigma\right)$ or $V_{\alpha}$. Here, we
choose $\eta_{\alpha}$ such that $\Phi_{\alpha}=N_{\alpha}/\widetilde{N}_{\alpha}$,
where $N_{\alpha}\left(\widetilde{N}_{\alpha}\right)$ is the number
of clusters in the structure that are symmetrically equivalent to
cluster $\alpha$, (without) taking into account the decorations.
This normalization gives $0\leq\Phi_{\alpha}\leq1$ for all $\alpha$'s,
which is convenient because the convergence of $V_{\alpha}$ with
respect to cluster size would directly reflect the convergence of
the CE. 

In practice, we use occupation variables $\xi$ to describe the atomic
species at each lattice site of a structure: $\xi_{A}\left(\sigma_{j}\right)$
equals 1 (0) if site $j$ in structure $\sigma$ is (not) occupied
by species $A\in\left\{ \text{Mo},\text{V},\text{Nb}\right\} $. Note
that this is distinct from the orthogonal basis in an alternate CE
formalism \citep{Sanchez1984}. Then, the correlation function of
structure $\sigma$ with respect to cluster $\alpha$ is computed
using

\begin{eqnarray}
\Phi_{\alpha}\left(\sigma\right) & = & \frac{1}{\widetilde{N}_{\alpha}}\sum_{c}\prod_{j\in c}\xi_{c_{j}}\left(\sigma_{j}\right),
\end{eqnarray}
where the sum is over all clusters $c$ symmetrically equivalent to
$\alpha$. The product is over all sites $j$ in the cluster, with
$c_{j}$ giving the atomic species at site $j$. We reiterate that
for ternary alloys and beyond, decorations need to be taken into account
when considering symmetrically equivalent clusters.

\subsection{Formation energy}

In general, either the configuration energy $E\left(\sigma\right)$
or the formation energy $E_{F}\left(\sigma\right)$ could be used
to train the CE. In this work, we use the latter, which is defined
as

\begin{eqnarray}
E_{F}\left(\sigma\right) & = & E\left(\sigma\right)-\sum_{A}\rho_{A}\left(\sigma\right)E\left(\sigma_{A}^{\text{pure}}\right),
\end{eqnarray}
where $\rho_{A}\left(\sigma\right)$ is the concentration of species
$A$ in the structure $\sigma$, and $\sigma_{A}^{\text{pure}}$ is
the pure system of species $A$. With the CE of $E\left(\sigma\right)$
from Eq. \ref{eq:clus-expansion-1}, the formation energy can be expanded
in terms of the ECIs: 
\begin{eqnarray}
E_{F}\left(\sigma\right) & = & \sum_{\alpha}\left[\Phi_{\alpha}\left(\sigma\right)-\sum_{A}\rho_{A}\left(\sigma\right)\Phi_{\alpha}\left(\sigma_{A}^{\text{pure}}\right)\right]V_{\alpha}.\label{eq:form-energy}
\end{eqnarray}
Because the expression in the square bracket vanishes exactly for
the empty cluster and singlets, the formation energy is expandable
in terms of just pairs and larger clusters \citep{Zarkevich2008}.
This form of the formation energy also naturally gives $E_{F}=0$
for pure systems. Then, writing Eq. \ref{eq:form-energy} as the linear
regression problem $y=X\beta$, we standardize the columns of $X$
to have unit $\ell_{2}$-norm before applying group lasso (or lasso),
as per common practice \citep{Hastie2015}. That is, denoting the
$i$th column of $X$ by $x_{i}$, we apply the invariant rescaling
$x_{i}\rightarrow x_{i}/\left\Vert x_{i}\right\Vert _{2}$ and $\beta_{i}\rightarrow\beta_{i}\left\Vert x_{i}\right\Vert _{2}$
such that $\left\Vert x_{i}\right\Vert _{2}=1$ for all $i$'s. 

\begin{figure}
\includegraphics[scale=0.5]{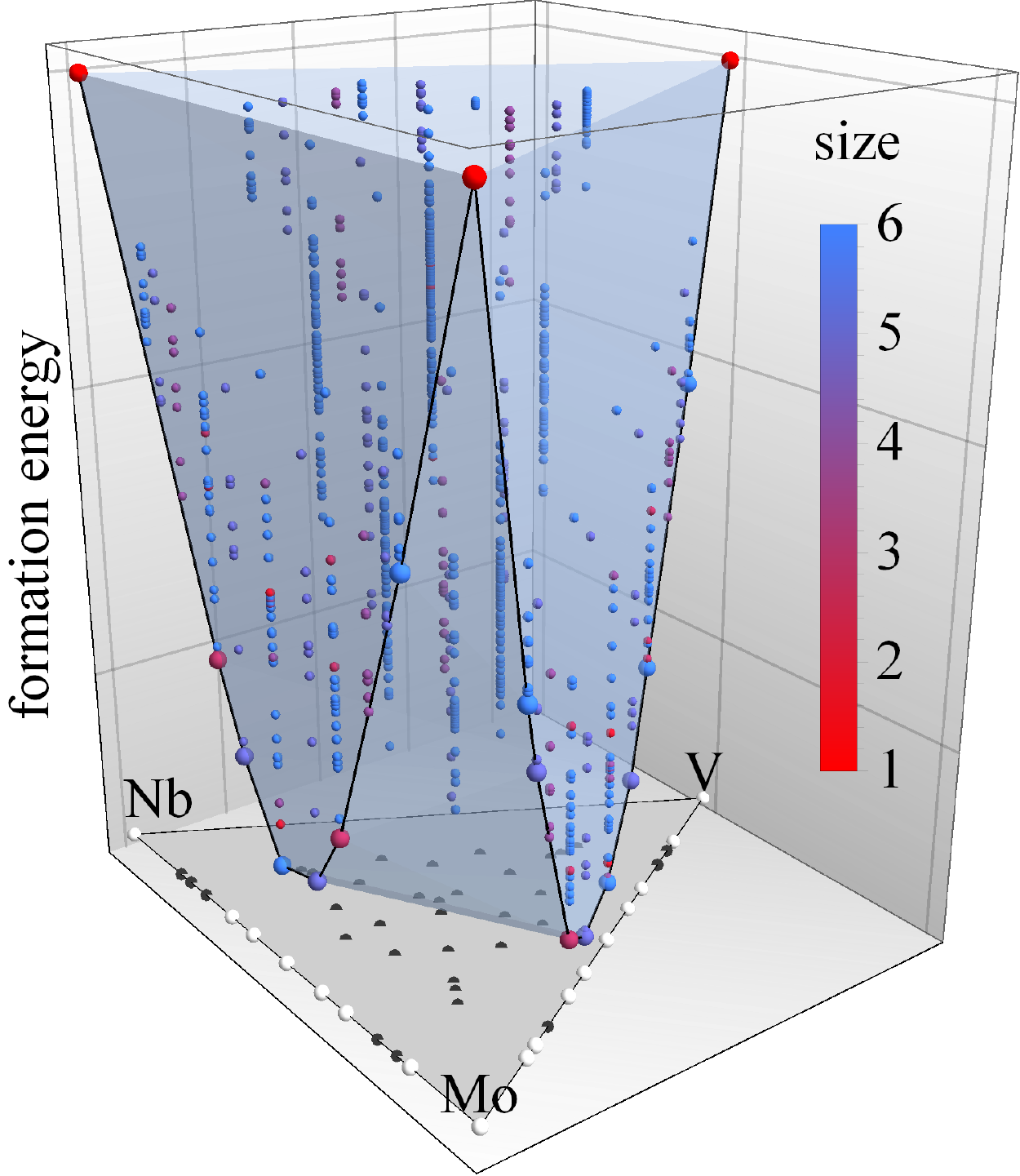}

\caption{The DFT formation energies $E_{F}$ of $1081$ derivative structures
with up to 6-atom unit cell in a bcc lattice, with respect to compositions.
Structures with $E_{F}>0$ are not shown. Redder (bluer) points are
structures with smaller (larger) unit cells. The blue translucent
surface is the ground state hull, with ground state structures represented
by larger points. The ternary plot shows the compositions of the structures,
with ground state structures highlighted in white. \label{fig:convex-hull-ternary-plot}}
\end{figure}

\subsection{Generation of training and test structures}

Ideally, the structures in a training set should be sufficiently varied
to capture all important physics of the system. To cover a wide range
of the configurational space, training structures can be selected
either randomly \citep{Nelson2013,Nelson2013a} or systematically
to maximize the covariance matrix of the correlation functions \citep{Seko2009}. 

In practice, computational constraints limit the number of DFT calculations
and favor training structures with smaller unit cells. This limitation
is especially severe for ternary alloys and beyond, because the configurational
space grows combinatorially with the number of atomic species. Therefore,
we select our training structures from a pool of 1081 derivative structures,
systematically generated up to 6-atom unit cell \citep{Hart2008,Hart2012}.
Fig. \ref{fig:convex-hull-ternary-plot} shows the DFT formation energies
and compositions of these structures. Notably, lower energy structures
tend to have smaller unit cells. Following the smallest-first algorithm
\citep{Ferreira1991}, structures with smaller unit cells are chosen
first. We exclude the three pure systems because their formation energies
given by Eq. \ref{eq:form-energy} are identically zero.

\begin{figure}
\includegraphics[scale=0.5]{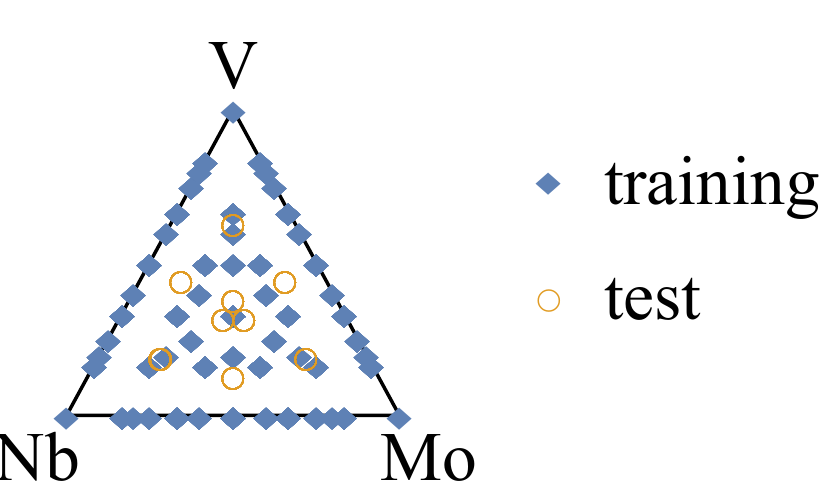}

\caption{A ternary plot showing the compositions of the 1081 training structures
and 500 test structures.\label{fig:composition-ternary-plot}}
\end{figure}

To verify that such training sets suffice for ternary systems, we
test the CE trained using small structures against a test set (holdout
set) of larger structures. The test set consists of 500 randomly selected
16-atom derivative structures; this set is not used to train our CE
model, but it serves to determine the testing/prediction error. The
ternary plot in Fig. \ref{fig:composition-ternary-plot} shows the
compositions of these test structures compared to those of the training
set.

\subsection{Initial set of clusters}

In our CE model, we treat V and Nb as the independent species, while
Mo is treated as dependent. As such, only clusters formed by Mo and
V atoms are required. In the bcc lattice, we consider up to the 9th-nearest-neighbor
(9NN) pairs, triplets with a 5NN cutoff, and four-body to six-body
clusters with a 3NN cutoff. These correspond to an initial pool of
239 symmetrically distinct clusters, consisting of 27 pairs, 84 triplets,
54 four-body clusters, 56 five-body clusters, and 18 six-body clusters.
Among these clusters are 1654 subcluster relations, which group lasso
uses to derive the final truncated CE based on the cluster selection
rules.

\begin{figure}
\includegraphics[scale=0.5]{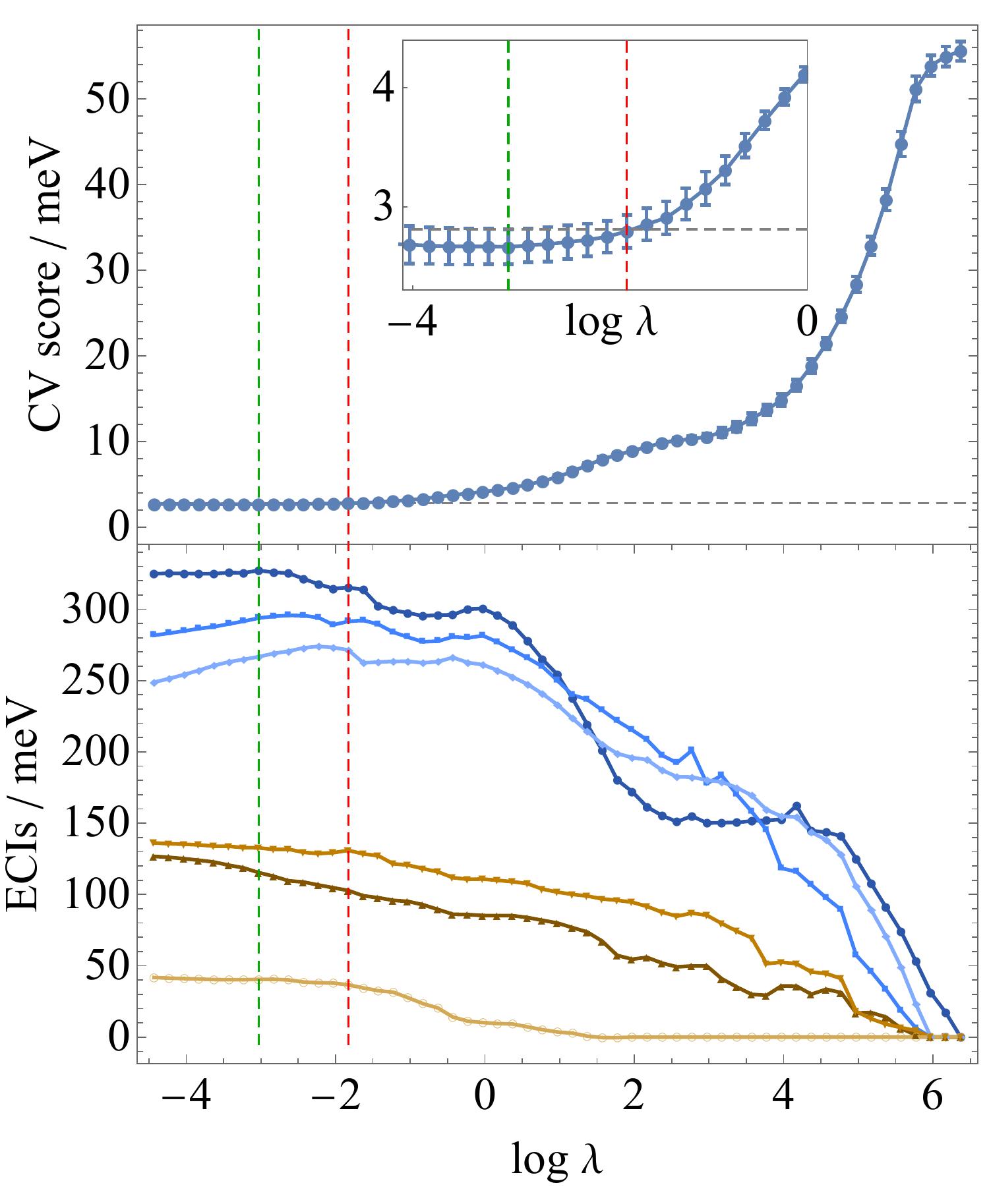} \caption{Top: The five-fold cross validation (CV) score of group lasso with
respect to the hyperparameter $\lambda$, using $800$ training structures.
The green vertical line denotes the minimum CV score. The red vertical
line is one-standard error away from the minimum and gives the optimal
$\lambda$. Inset: a closeup of the same plot. Bottom: The six ECIs
of the nearest-neighbor (blue) and next-nearest-neighbor (yellow)
pairs with respect to $\lambda$. \label{fig:cv-score-vs-lambda}}
\end{figure}

\subsection{Tuning of hyperparameter $\lambda$}

Using the DFT formation energies of the training structures, we use
group lasso to select a properly truncated CE set from the initial
239 distinct clusters. The group lasso minimization problem is efficiently
solved using a block coordinate descent algorithm \citep{Hastie2015},
which reduces the multidimensional minimization problem to a sequence
of root-finding problems in 1D. Overfitting (underfitting) happens
when the hyperparameter $\lambda$ is too small (large). The optimal
$\lambda$ is selected based on a five-fold cross validation (CV)
with the one-standard error rule \citep{Hastie2015}, as illustrated
in Fig. \ref{fig:cv-score-vs-lambda} (top). I.e., the optimal $\lambda$
corresponds to the most regularized model with CV score within one
standard error of the minimal CV score. The bottom plot of Fig. \ref{fig:cv-score-vs-lambda}
shows coefficient shrinkage and cluster selection in group lasso.
As $\lambda$ decreases, the model becomes less regularized and the
ECIs generally increase; the solution is also less sparse as more
ECIs become nonzero.

\bibliographystyle{apsrev4-1}
\bibliography{library}

\end{document}